\documentstyle[aps,preprint,prl,epsfig]{revtex}
\tightenlines
\begin{document}

\draft
\title{SU(3) realization of the rigid asymmetric rotor within the IBM}
\author{Yuri F.~Smirnov$^a$,
        Nadya A.~Smirnova$^b$
        and Piet Van~Isacker$^b$}
\address{$^a$ Instituto de Ciencias Nucleares,
              UNAM, M\'exico DF, 04510, M\'exico} 
\address{$^b$ Grand Acc\'el\'erateur National d'Ions Lourds,
              BP 5027, F-14076 Caen Cedex 5, France}
\date{\today}
\maketitle

\begin{abstract}
It is shown that the spectrum of the asymmetric rotor
can be realized quantum mechanically
in terms of a system of interacting bosons.
This is achieved in the SU(3) limit
of the interacting boson model
by considering higher-order interactions between the bosons.
The spectrum corresponds to that of a rigid asymmetric rotor
in the limit of infinite boson number.
\end{abstract} 


It is well known
that the dynamical symmetry limits
of the simplest version of the interacting boson model
(IBM)~\cite{ArIa75,IaAr87}, IBM-1,
correspond to particular types of collective nuclear spectra.
A Hamiltonian with U(5) dynamical symmetry~\cite{ArIa76}
has the spectrum of an anharmonic vibrator,
the SU(3) Hamiltonian~\cite{ArIa78}
has the rotation-vibration spectrum
of vibrations around an axially symmetric shape
and  the SO(6) Hamiltonian~\cite{ArIa79}
yields the spectrum of a $\gamma$-unstable nucleus~\cite{Wilets}.
There exists another interesting type of spectrum
frequently used to interpret nuclear collective excitations
which corresponds to the rotation of a rigid asymmetric top~\cite{DaFi58}
and which, up to now, has found no realization
in the context of the IBM-1.
The purpose of this letter
is to extend the IBM-1 towards high-order terms
such that a realization of the rigid non-axial rotor
of Davydov and Filippov becomes possible.
A pure group-theoretical approach is used
that allows to establish the connection
between algebraic and geometric Hamiltonians
not only from the comparison of their spectra
but also from the underlying group properties.

Let us first recall some of the aspects
that have enabled a geometric understanding of the IBM.
The relation between the Bohr-Mottelson collective model~\cite{BM}
and the IBM has been established~\cite{GiKi80,DiSc80}
on the basis of an intrinsic (or coherent) state for the IBM.
Via this coherent-state formalism,
a potential energy surface $E(\beta,\gamma)$
in the quadrupole deformation variables $\beta$ and $\gamma$
can be derived for any IBM Hamiltonian
and the equilibrium deformation parameters $\beta_0$ and $\gamma_0$
are then found by minimizing $E(\beta,\gamma)$.
It is by now well established
that a one- and two-body IBM-1 Hamiltonian
can give rise only to axially symmetric equilibrium shapes 
($\gamma_0=0^{\rm o} \; {\rm or} \; 60^{\rm o}$)~\cite{GiKi80,DiSc80}
and that a triaxial minimum in the potential energy surface  
requires at least three-body interactions~\cite{PVICh81}.  


Since the relationship between $\gamma $-unstable model and rigid triaxial 
rotor was always an open question, Otsuka {\it et al.}~\cite{OtSu87,SuOt89}
investigated in detail the SO(6) solutions of one- and two-body IBM-1 
Hamiltonian.
They found out that the triaxial intrinsic state with $\gamma_0=30^{\rm o}$ 
produces after the angular momentum projection 
the exact SO(6) eigenfunctions for small numbers of bosons $N$.
Thus they conclude that for finite boson systems triaxiality
reduces to $\gamma $-unstability.


If three-body terms are included in the IBM-1 Hamiltonian
a triaxial minimum of the nuclear potential energy surface
can be found
and numerous studies of the corresponding spectra
have been performed~\cite{PVICh81,HePVI84,CaBr85,LeSh90,ZaCa91,LJ94,LJ95}.
However, the existence of a minimum in the potential energy surface
at $0^{\rm o}<\gamma<60^{\rm o}$
is not a sufficient condition for a rigid triaxial shape
since this minimum can be shallow
indicating a $\gamma$-soft nucleus.
The interest in higher-order terms in IBM-1
has been renewed recently 
by the challenging problem of anharmonicities
of double-phonon excitations
in well-deformed nuclei~\cite{GRPVI99,PVI99}. 
Although their microscopic origin is not clear at the moment,
the occurrence of higher-order interactions
can be understood qualitatively
as a result of the projection of two-body interactions
in the proton-neutron IBM~\cite{ArOt77}
onto the symmetric IBM-1 subspace.
For example, it is well known 
that triaxial deformation arises
within the SU$^*$(3) dynamical symmetry limit~\cite{Dieperink}
of the proton-neutron IBM
without any recourse to interactions
of order higher than two.

Nuclear collective states are treated in IBM-1
in terms of $N$ bosons of two types:
monopole ($l^\pi=0^+$) $s$ bosons
and quadrupole ($l^\pi=2^+$) $d$ bosons~\cite{IaAr87}.
For a given nucleus,
$N$ is the half number of valence nucleons (or holes)
and is thus fixed. 
Analytical solutions can be constructed
for particular forms of the Hamiltonian
which correspond to one of the three possible reduction chains
of the dynamical group of the model U(6):
\begin{equation}
\label{DS}
{\renewcommand{\arraystretch}{0.1}
\begin{array}{ccl}
& & \mbox{U(5)}\supset \mbox{SO(5)}\supset
\mbox{SO(3)}  \\
& \nearrow   \\
\mbox{U(6)} & \rightarrow & \mbox{SU(3)}\supset \mbox{SO(3)} \\
& \searrow &  \\
& & \mbox{SO(6)}\supset \mbox{SO(5)}\supset \mbox{SO(3)} \\
\end{array} }.
\end{equation}  
The SU(3) dynamical symmetry Hamiltonian
corresponds to a rotation-vibration spectrum
of a vibrations around an axially symmetric shape
which in the limit $N\to\infty$
goes over into the spectrum of a rigid axial rotor~\cite{GiKi80}.
The two other reduction chains in~(\ref{DS})
contain the SO(5) group
whose Casimir invariant exactly
corresponds to $\gamma$-independent potential
of Wilets and Jean~\cite{Wilets}
and is responsible for $\gamma$-soft character
of the spectrum~\cite{DBes,LeNo86}.

To obtain a rigid (at least for $N\to\infty$) triaxial rotor
the starting point is the SU(3) limit of the IBM-1
with higher-order terms in the Hamiltonian.
This approach is inspired by Elliott's SU(3) model~\cite{Elliott,EllHa63}
where the rotor dynamics is well established
for SU(3) irreducible representations (irreps)
with large dimensions.
 
Following Ref.~\cite{PVI85},
we consider the most general SU(3) dynamical symmetry Hamiltonian
constructed from the second, third and fourth order 
invariant operators of the SU(3) $\supset$ SO(3) integrity basis~\cite{JuMi74}:
\begin{equation}
\label{H}
H_{\rm IBM}=H_0+aC_2+b C_2^2+cC_3+d\Omega +e\Lambda +f L^2+g L^4+h C_2 L^2 \;. 
\end{equation}    
Here the following notation is used:
\begin{eqnarray}
\label{C2}
& &\begin{array}{r@{\ }l}
C_2&\equiv C_2[\mbox{SU(3)}]=2 Q^2 + \frac34 L^2 \; ,
\end{array} 
\\
\label{C3}
& &\begin{array}{r@{\ }l}
C_3&\equiv C_3[\mbox{SU(3)}]
=-\frac49 \sqrt{35}[Q\times Q \times Q]^{(0)}_0
-\frac{\sqrt{15}}{2}[L\times Q \times L]^{(0)}_0  \; , \\
\end{array}
\\
\label{Omega}
& &\begin{array}{r@{\ }l}
\Omega &=-3 \sqrt{\frac52} [L\times Q\times L]^{(0)}_0 \; , \\
\end{array}
\\
\label{Lambda}
& &\begin{array}{r@{\ }l}
\Lambda &=\left[L \times Q \times Q \times L \right]^{(0)}_0 \; , \\
\end{array}
\end{eqnarray}
where
\begin{eqnarray}
&&L_q=\sqrt{10}[d^+\,{\times}\, \tilde d]^{(1)}_q\; , \label{L} \\
&&Q_q=[d^+\,{\times}\,s+s^+\,{\times}\, \tilde d]^{(2)}_q
-\frac{\sqrt{7}}{2}[d^+\,{\times}\, \tilde d]^{(2)}_q \; , \label{Q}
\end{eqnarray}
are SU(3) generators, satisfying the standard commutation relations,
\begin{equation}
\label{comSU3}
\begin{array}{l}
[L_q,L_{q'}]=
-\sqrt{2}(1q 1 q'|1 q + q') L_{q +q'}\; , \\[0pt]
[L_q,Q_{q'}]=
-\sqrt{6}(1 q 2 q'|2 q +q ') Q_{q +q'}\; , \\[0pt]
[Q_q,Q_{q'}]=
\frac34\sqrt{\frac52}(2 q 2 q'|1 q +q')L_{q + q'} \; .
\end{array}
\end{equation}      

In the context of the shell model,
an SU(3) Hamiltonian of the type~(\ref{H}) 
has been considered by a number of 
authors~\cite{DrRo85,LeDr86,Carv86,CaDr88,Roch88,BaEs95}.
Specifically, it was established~\cite{LeDr86}
that the rotor Hamiltonian can be constructed from $L^2$
and the SU(3) invariants $\Omega$ and $\Lambda$.
This follows from the asymptotic properties
of these SU(3) invariants
whose spectra do correspond to rigid triaxial rotor 
for SU(3) irrep labels $\lambda,\mu\to\infty $.
In addition, a relation between ($\lambda,\mu$)
and the collective variables ($\beta,\gamma$)
characterizing the shape of the rotor
can be derived~\cite{CaDr88}.
In contrast, all attempts so far to construct a rigid rotor
in the SU(3) dynamical symmetry limit of the IBM-1,
even if including higher-order terms, 
are restricted to axial shapes~\cite{PVI85}.

A noteworthy difference should be pointed out
between the SU(3) realizations in the shell model and the IBM
and concerns the irreps that occur lowest in energy.
In the shell model the ground-state irrep is dictated by the
leading shell-model configuration. 
For example, it is (8,4) for $^{24}$Mg
and (30,8) for $^{168}$Er~\cite{Elliott,LeDr86}.
In the IBM the lowest representation
is determined by the Hamiltonian.
One can show that for an SU(3) Hamiltonian
with two- and three-body interactions
it is either $(2N,0)$ or $(0,N)$,
which corresponds to axially symmetric nucleus.
The essential point that is exploited here
is that this choice of the lowest SU(3) irrep
becomes more general for the IBM-1 Hamiltonian
with up to four-body terms.

One can show that with the linear combination
\begin{equation}
\label{LC}
F=a C_2 + b C_2^2 \; ,
\end{equation}
any given irrep $(\lambda_0,\mu_0)$
that occurs for a system of $s$ and $d$ bosons
can, in principle, be brought lowest in energy.
The proof is as follows.
The eigenvalue of the second-order SU(3) Casimir invariant is
\begin{equation}
\label{g2}
g_2(\lambda ,\mu )=\lambda^2 + \mu^2 +\lambda \mu +3\lambda +3 \mu \; .
\end{equation}
Within a given U(6) irrep $[N]$
the following SU(3) $(\lambda,\mu)$ values
are admissible~\cite{Elliott}:
\begin{equation}
\begin{array}{ll}
(\lambda ,\mu )=&(2N,0),\; (2N-4,2), \; \ldots \; 
(2,N-1) \quad {\rm or} \quad (0,N) \; ,\\[0pt]    
&(2N-6,0),\; (2N-10,2), \; \ldots \; 
(2,N-4) \quad {\rm } \quad (0,N-3) \; ,\\[0pt]   
& \ldots \\
&(4,0), \; (0,2)\quad {\rm for} \quad N({\rm mod}3)=2 \; ,\\[0pt]
& (2,0) \quad {\rm for} \quad N({\rm mod}3)=1 \; ,\\[0pt]
& (0,0) \quad {\rm for} \quad N({\rm mod}3)=0 \; .\\[0pt]   
\end{array}
\end{equation} 
For the first row in this equation
(which corresponds to the largest values of Casimir invariants)
one has the relation 
\begin{equation}
\label{Nlm}
2N=\lambda + 2 \mu  \; ,
\end{equation}
and consequently $g_2(\lambda,\mu )\equiv g_N(\lambda)$.
The eigenvalues of (\ref{LC}) are thus given by
\begin{equation}
\label{eg2}
F_N(\lambda )=a g_N(\lambda )+ b g_N^2(\lambda ) \; .
\end{equation}
Minimization of this function with respect to $\lambda$
gives the following condition
for the irrep $(\lambda_0,\mu_0)$ to be lowest in energy:
\begin{equation}
\label{param}
\frac{a}{b} =-2 g(\lambda _0, \mu _0) \; .
\end{equation}
It is worth mentioning
that one fails to obtain a similar minimization condition
with just the second- and third-order SU(3) Casimir invariants.

Once $(\lambda_0,\mu_0)$ is fixed,
the constant part of the Hamiltonian~(\ref{H})
takes the form
\begin{equation}
\label{H0}
\tilde H_0=H_0- b g_2^2(\lambda_0,\mu_0 )+c g_3(\lambda _0,\mu_0 )\;, 
\end{equation}    
where $g_3(\lambda,\mu)$ is the expectation value of $C_3$[SU(3)]
\begin{equation}
\label{g3}
g_3(\lambda ,\mu )=
\frac19(\lambda -\mu )(2\lambda + \mu +3)(\lambda + 2 \mu +3)\; ,
\end{equation}
and the Hamiltonian~(\ref{H}) can be rewritten as
\begin{equation}
\label{HF}
H_{\rm IBM}=\tilde H_0 + d \Omega + e \Lambda +\tilde f L^2+ g L^4 \;, 
\end{equation}  
where $\tilde f = f + h g_2(\lambda_0,\mu_0)$.

To see the relation between this Hamiltonian 
and that of the rotor,
we rewrite (\ref{HF}) further as
\begin{equation}
\label{HFa}
H_{\rm IBM}=\tilde H_0 + d (L_i Q_{ij} L_j) + e (L_i Q_{ik} Q_{kj}L_j) 
+\tilde f L^2+ g L^4 \; ,                       
\end{equation}  
where $Q_{ij}$ are cartesian components
of the quadrupole tensor~(\ref{Q}).
This operator has the same functional form
as the rigid rotor Hamiltonian
\begin{equation}
\label{rotor}
H_{\rm rotor}= A_i {\rm L}_i^2 = (L_i M_{ij} L_j) \;, 
\end{equation}  
where ${\rm L}_i$ ($L_i$) are the components of the angular momentum $L$
in the intrinsic (laboratory) frame,
$M_{ij}$ is a moments of inertia tensor with the constant components.
The parameters of inertia $A_i$ are related 
to the principal moments of inertia ${\cal I}_i$ as
\begin{equation}
\label{mi}
A_i=\frac{\hbar^2}{2{\cal I}_i}  \; .
\end{equation} 
The rotor moments of inertia tensor
is connected with the SU(3) quadrupole tensor components through
\begin{equation}
\label{it}
M_{ij}=[\tilde f + g L(L+1)] \delta_{ij} + d Q_{ij} + e Q_{ik}Q_{kj} \; .
\end{equation} 
Note that, in general,
the components of the quadrupole tensor $Q_{ij}$
fluctuate around their average values. 
When these fluctuations are negligible,
the spectrum of the IBM-1 Hamiltonian (\ref{HFa})
is close to the spectrum of rigid asymmetric rotor
with corresponding moments of inertia.

The analogy between the rigid rotor
and SU(3) dynamical symmetry Hamiltonians
can be studied from a group-theoretical point of view.
The dynamical group of the quantum rotor~\cite{Ui70}
is the semidirect product T$_5$ $\wedge$ SO(3)
where T$_5$ is generated by the five components of
the collective quadrupole operator $\overline Q$. 
The operators $\overline Q$ and $L$ satisfy the commutation relations
\begin{equation}
\label{comRotor}
\begin{array}{l}
[L_q,L_{q'}]=-\sqrt{2}(1q 1 q'|1 q+q')L_{q+ q'}\;,\\[0pt]
[L_q,\overline Q_{q'}]=
-\sqrt{6}(1q 2 q'|2 q+q') \overline Q_{q+ q'}\; , \\[0pt]
[\overline Q_q,\overline Q_{q'}]=0 \; ,
\end{array}
\end{equation}
which define the rotor Lie algebra t$_5$ $\oplus$ so(3).
The only difference between (\ref{comSU3}) and (\ref{comRotor})
is in the last commutator.

The replacement $Q \to Q/\sqrt{C_2}$ in the su(3) algebra leads to
$[Q_q,Q_{q'}]\to 0$ for $\lambda, \mu \gg L$.
Thus for large $\lambda $ and $\mu $
the su(3) algebra contracts to
the rigid rotor algebra t$_5$ $\oplus$ so(3).
The irreps of t$_5$ $\oplus$ so(3)
are characterized by the $\beta$ and $\gamma$ shape variables
which can be related to SU(3) irrep labels $\lambda$ and $\mu$
as in Ref.~\cite{CaDr88}:
\begin{equation}
\label{rel}
\begin{array}{l}
\kappa \beta \cos{\gamma }= \frac13 (2 \lambda + \mu + 3) \; , \\
\kappa \beta \sin{\gamma }= \frac{1}{\sqrt{3}} (\mu + 1) \; ,\\
\end{array}
\end{equation}
where $\kappa$ has to be determined from parameterization
\begin{equation}
\label{QRotor}
\overline Q_q=\kappa \beta [\delta_{q 0} \cos{\gamma }
+ \frac{1}{\sqrt{2}} (\delta_{q 2} + \delta_{q, -2}) \sin{\gamma }] \;,
\end{equation}
with $\beta \ge 0$ and $0^{\rm o} \le \gamma \le 60^{\rm o}$.

The difference between the shell model SU(3) realization
and the SU(3) dynamical symmetry limit of the IBM
can be visualized on a $(\beta,\gamma)$ plot
which gives the relation between $(\beta,\gamma)$
and the SU(3) labels $(\lambda,\mu)$~\cite{CaDr88} (see Figure~1).
The SU(3) irreps valid for the IBM (marked by circles) are only a subset
of those which are allowed in the shell model in accordance with 
the Pauli principle (e.g., Fig.~2 in Ref.~\cite{BaEs95}).

From a group-theoretical point of view,
the difference is seen from the following considerations.
The invariant symmetry group of the asymmetric rotor
is the point symmetry group D$_2$,
whose irreps can be classified as $A_1$, $B_1$, $B_2$ and $B_3$. 
In the contraction limit,
the $(\lambda,\mu)$ irrep of SU(3)
reduces to one of the D$_2$ irreps
according to the even or odd values of $\lambda$ or $\mu$. 
Since the IBM only allows even $\lambda$ and $\mu$,
only totally symmetric $A_1$ levels of the asymmetric rotor
can be represented. 
These are also the asymmetric rotor levels, 
which should be considered~\cite{DaFi58} 
in connection with nuclear spectra.  

As an example, in Figure~2
the spectrum of the Hamiltonian~(\ref{H}) is shown
with the parameters 
$a=-72$ keV,
$b=0.1$ keV,
$d=0.1$ keV,
$f=25$ keV,
$c=e=g=h=0$
and $H_0=12960$ keV
for the two lowest SU(3) irreps $(10,10)$ and $(14,8)$
of an $N=15$ boson system
and compared with the spectrum
of an asymmetric rotor of highest asymmetry 
($\gamma=30^{\rm o}$).  
The matrices of the operators $\Omega $ and $\Lambda $ in the Elliott's basis
can be found in Refs.~\cite{PVI85,FiOvSm}.
The SU(3) spectrum consists of $(\lambda,\mu)$-multiplets
within which the levels are arranged in bands characterized 
by Elliott's quantum number $K$ where for even $\lambda $ and $\mu $
\begin{equation}
\label{KL}
\begin{array}{l}
K={\rm min}\{\lambda, \mu \}, {\rm min}\{\lambda, \mu \}-2 , \; \ldots \; , 
2 \; {\rm or} \; 0 \; ,\\
L=K, K+1, \; \ldots \; , {\rm max}\{\lambda, \mu \} \quad {\rm for} \; 
K\ne 0 \; ,\\ 
L=0, 2, \; \ldots \; , {\rm max}\{\lambda, \mu \} \quad {\rm for} \; K=0 \;.\\ 
\end{array}
\end{equation}
The spectrum of a triaxial rotor
consists of a ground-state band with $L=0,2,4, \; \ldots $
and an infinite number of the so-called abnormal bands:
$L=2,3,4,5,\; \ldots $ (1st abnormal band),    
$L=4,5,6,7,\; \ldots $ (2nd abnormal band), and so on. 
Contrary to the axially symmetric case,
the projection of the angular momentum $L$
on the intrinsic $z$-axis no longer is a good quantum number.
Also, the spectrum contains $L/2+1$ states for $L$ even
and $(L-1)/2$ states for $L$ odd.
It is seen that the low-energy spectrum
corresponding to $(10,10)$ irrep
is remarkably close to the spectrum of the asymmetric rotor.
The resemblance includes the prominent even-odd staggering
in the first abnormal band
which is a perfect signature
to distinguish axial, rigid or soft triaxial rotors
(see, e.g. Refs.~\cite{ZaCa91,LJ95}).
Eventual differences between the triaxial rotor
and the SU(3) calculation
are caused by the finite number of allowed $L$-values
for each $K$ in a given $(\lambda,\mu)$ irrep.

Although the above considerations
were limited to the SU(3) dynamical symmetry, 
we would like to stress
that this is not the only possible realization 
of rigid triaxiality in the IBM-1.
As has been demonstrated recently~\cite{PVI99},
a rotational spectrum
can be generated by only quadratic $[Q^0 \times Q^0]^{(0)}$
and cubic $[Q^0 \times Q^0 \times Q^0]^{(0)}$
SO(6) invariants where 
\begin{equation}
\label{Q0}
Q^0_q=[d^+\,{\times}\,s+s^+\,{\times}\, \tilde d]^{(2)}_q
\end{equation}  
is an SO(6) generator. 
This can be understood from a group-theoretical point of view.
In the limit $[Q^0_q,Q^0_{q'}] \to 0$
the so(6) algebra contracts
to the rigid rotor algebra t$_5$ $\oplus$ so(3)
and a rigid rotor realization with
SO(6) dynamical symmetry is obtained in IBM-1.

Inspired by this result,
it would be of interest to inspect the Hamiltonian 
of the type~(\ref{H}) 
\begin{equation}
\label{HG}
H_{\rm IBM}=H_0+ \alpha (L_i Q^{\chi }_{ij} L_j) 
+ \eta (L_i Q^{\chi }_{ik} Q^{\chi }_{kj}L_j) 
+\zeta L^2+ \xi L^4 \; ,                       
\end{equation}
where $Q^{\chi }$ is a general quadrupole operator
\begin{equation}
\label{QCHI}
Q^{\chi }_q=[d^+\,{\times}\,s+s^+\,{\times}\, \tilde d]^{(2)}_q
+ \chi [d^+\, {\times}\, \tilde d]^{(2)}_q \; .
\end{equation}  
One expects that,
provided the commutation relations for $L$ and $Q^\chi$ 
are close to those for the rotor~(\ref{comRotor}),
the spectrum of the Hamiltonian~(\ref{HG})
should resemble that of the rigid rotor.


In summary, an IBM-1 realization
of the rigid rotor has been suggested.
Although the example has been restricted
to the SU(3) dynamical symmetry, 
any Hamiltonian of a similar type
constructed with general quadrupole operator
and with the angular momentum operator
produces a rigid rotor spectrum 
under the condition of appropriate commutation relations.
The required and sufficient condition
to obtain the rotor dynamics
is the contraction of the dynamical algebra of the Hamiltonian
to the rigid rotor algebra t$_5$ $\oplus$ so(3).

We gratefully acknowledge discussions with
P.~von Brentano, 
F.~Iachello,
D.~F.~Kusnezov,
S.~Kuyucak,
A.~Leviatan,
N.~Pietralla,
A.~A.~Seregin
and A.~M.~Shirokov.
N.A.S.\ and P.V.I.\ thank the Institute for Nuclear Theory
at the University of Washington for its hospitality and support.
Yu.F.S.\ thanks GANIL for its hospitality
The work is partly supported by Russian Foundation of Basic Research 
(grant 99-01-0163).


\begin{figure}
\label{figure1}
 \epsfig{file=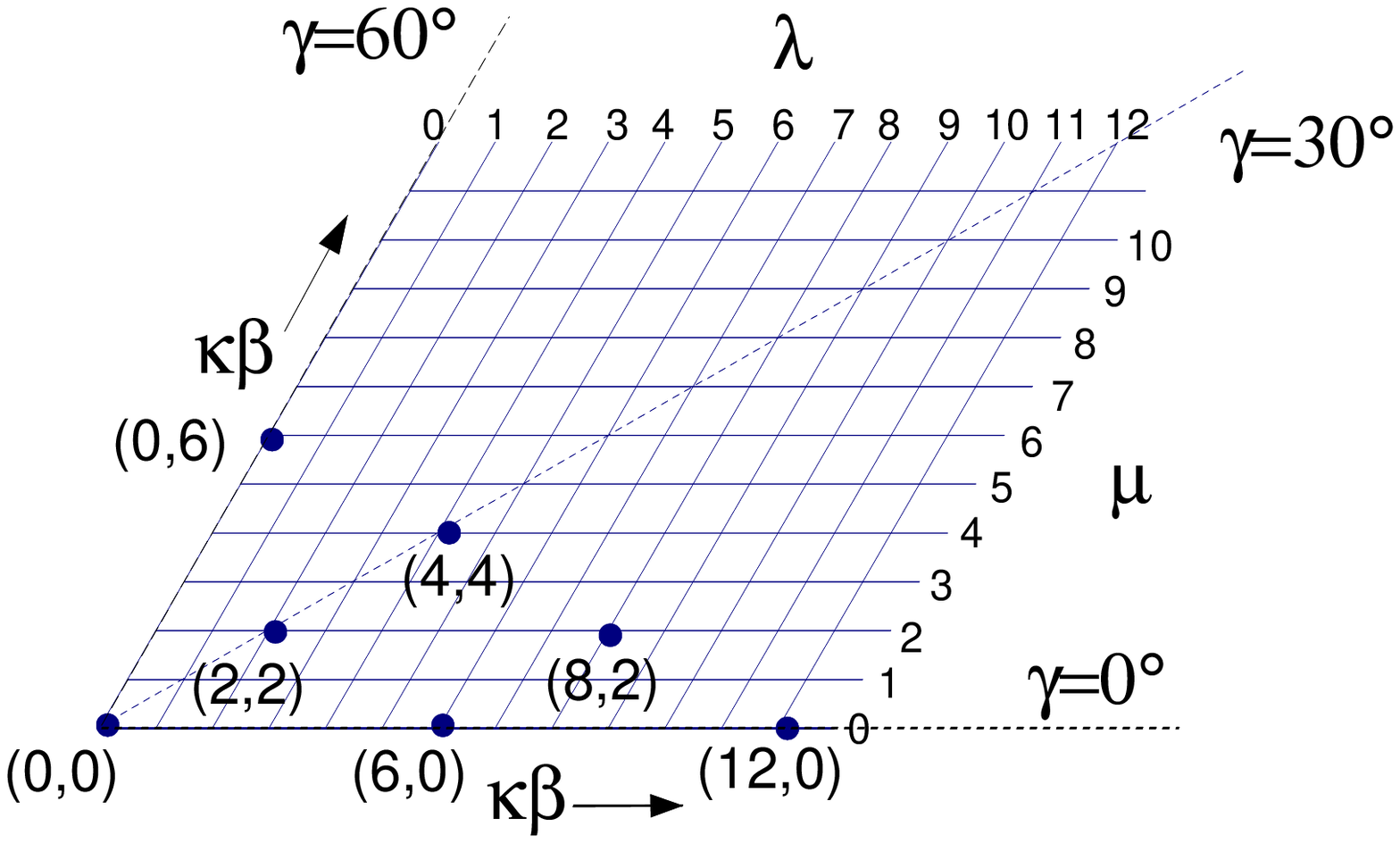,width=0.98\textwidth}
\caption{Relation between the collective rotor parameters $(\beta,\gamma)$
and the SU(3) irrep labels $(\lambda,\mu)$.
The circles indicate the irreps valid for the IBM with $N=6$ bosons.}
\end{figure} 
           
\begin{figure}
\label{figure2}
\centerline{\epsfig{file=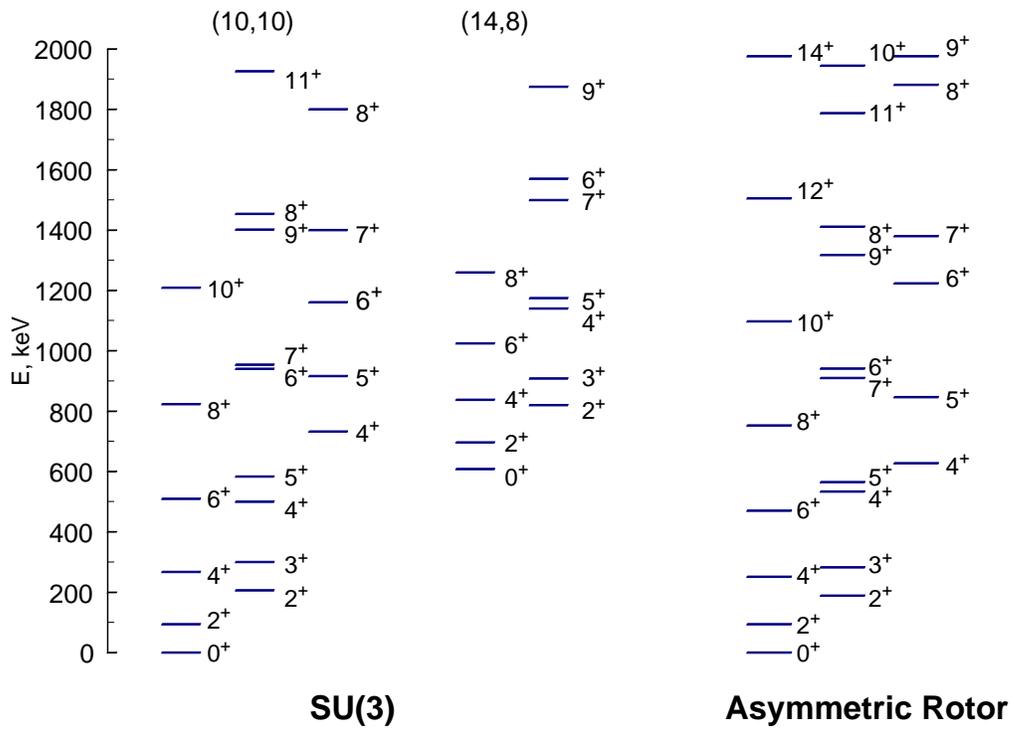,width=0.98\textwidth}}  
\caption{Left: a typical spectrum
of the SU(3) dynamical symmetry Hamiltonian~(\protect\ref{H})
with the set of parameters 
$a=-72$ keV,
$b=0.1$ keV,
$d=0.1$ keV,
$f=25$ keV,
$c=e=g=h=0$
and $H_0=12960$ keV for $N=15$ bosons.
The two lowest SU(3) irreps $(10,10)$ and $(14,8)$ are shown. 
Right: the asymmetric rotor spectrum for $\gamma =30^{\rm o}$.}
\end{figure} 
           
\end{document}